\documentclass{article}

\usepackage{graphicx}
\begin{document}
\title{Generalized Entropy with Clustering and Quantum Entangled States}

\author{F. Shafee \\Department of Physics, Princeton University,\\ Princeton, NJ 06540, USA}

\maketitle

\begin{abstract}
We first show how a new definition of entropy, which is intuitively very simple, as a divergence in cluster-size space, leads to a generalized form that is nonextensive for correlated units, but coincides exactly with the conventional one for completely independent units. We comment on the relevance of such an approach for variable-size microsystems such as in a liquid. We then indicate how the entanglement and purity of a two-unit compound state can depend on their entanglement with the environment. We consider
entropies of Tsallis, which is used in many different real-life contexts, and also our new generalization, which takes into account correlated clustering in a more transparent way, and is just as amenable mathematically as that of Tsallis,   and show how both purity and entanglement can appear naturally together in a measure of mutual information in such a generalized picture of the entropy, with values differing from the Shannon type of entropy. This opens up the possibility of using such an entropy in a quantum context for relevant systems, where interactions between microsystems makes clustering and correlations a non-ignorable characteristic.
\end{abstract}


\section{Introduction}

A new form of generalized nonextensive entropy recently proposed by
us\cite{FS1} has been shown  in  to give small but
interesting departures from the Shannon case in terms of
thermodynamic properties of a system in a manner similar to but also
somewhat different from Tsallis entropy \cite{TS1,TS2,PA1}.
Conceptually, the new entropy appears from a novel definition of
entropy in terms of the rescaled phase cells due to correlated clusters, and in a limit similar to the Tsallis case approaches Shannon's classical extensive entropy. Kaniadakis has also suggested \cite{KA1,KA2} the use of deformed functions leading
to unusual forms of entropy, from a kinetic Principle related to
phase space, which gives excellent results for cosmic ray spectra. In our case the definition of entropy is particularly simple, as it can be expressed simply as the divergence of a vector representing the modified probabilities for the different possible states taking into account a rescaling due to correlations or clustering due to interactions between the microsystems.

In microscopic systems quantum entanglement of states is also a relevant issue. Some authors \cite{VI1,AB1,AB2} have studied the problem of quantum
entanglement of two states in the picture of Tsallis type
nonextensive entropy. The generalization of Shannon entropy to the
very similar von Neumann entropy using density operators in place of
probability distributions \cite{NC1} reveals common features of the
stochastic and the quantum forms of uncertainties and this treatment
can be extended to Tsallis' form too.

Our purpose here is to present a combined study of stochasticity and
quantum entanglement, so that the former emerges from the quantum
picture in a natural way, and then we intend to show that our new
approach of defining entropy also allows us to obtain a measure of
mutual information that involves stochasticity and entanglement
together in a clear comprehensible way. The fact that our new definition of entropy, which is conceptually very simple, also gives the probability distribution function in a closed form in terms of Lambert $W$ functions \cite{FS1} allows one to carry out many calculations with the same ease as for Tsallis entropy. In this work, however, the probability distribution will not be needed for explicit use.

\section{A Generalized Form of Entropy from a New Definition}

Entropy is intuitively associated with randomness, because it is a measure of the loss of information about a system, or the indeterminacy of its exact state, which in turn depends on the probability distribution for various states. A uniform probability distribution function (pdf) among all states indicates maximal uncertainty in state space and gives the maximal entropy, whereas Dirac/Kronecker delta (continuous or discrete states) pdf with no uncertainty has zero entropy. Combinatorics gives the Boltzmann form

\begin{equation}
S = k \log [ N!/ \prod_i n_i!]
\end{equation}

because in equilibrium the $n_i$ are simply $ N p_i$, where $N$ is the total number of subsystems, and $n_i$ is the number of subsystems in the i-th state. In terms of the $p_i$ themselves one gets the Shannon form given below. It is well-known that maximizing the entropy with the constraints $ \sum p_i=1$   and (with $E_i$ the energy of the i-th state, and U the total energy, which is fixed) $\sum  p_i E_i= U$ gives the exponential probability distribution

\begin{equation}
p_i = C  \exp(- \beta E_i)
\end{equation}

where the Lagrange multiplier constant $\beta$ can be identified as the inverse of the temperature.

Let us now consider Shannon coding theorem \cite{NC1}: when the letters $A_i$  of the alphabet used in a code have the probabilities $p_i$, then it can be shown fairly easily that a stream of random letters coming out of the source with the given probabilities in the long run will relate the entropy to the probability of the given sequence:

\begin{equation}
P(sequence) = \prod_i p_i^{N p_i}= \exp[-N S]
\end{equation}

where $S$ is the entropy per unit and $N$ is the large number of letters in the sequence.

We shall now define entropy from a somewhat different viewpoint which takes into account interaction among the units, producing clusters of size $q_i$ units for the i-th state. This effective size may in general be a fraction, and if the interaction is weak,  the average cluster size $q_i$ is just over unity. If we think of liquid clusters, the typical subsystem in state $i$  may bean assembly of $r_i$ molecules, but this may change due to environmental factors, such as pH value, to $s_i$, so that we have a rescaling value of $q_i= s_i/r_i$, which may be greater or less than 1.  In general we allow $q_i$ to be different for each $i$. Since $p_i$ is the probability of a single occurrence of the i-th state, i.e. for a cluster of size unity (which may consist of a typical number of subunits),  the probability for the formation of a cluster of size $q_i$ is $p(q_i) = (p_i)^{q_i}$. Let us now consider the vector

\begin{equation}
v_i = p(q_i) = (p_i)^{q_i}
\end{equation}

This is $n$-dimensional, where $n$ is the number of single unit states available to the system components.

Let us now consider the ``phase space" defined by the $q_i$ co-ordinates. As we have said above, the deviations of these parameters from unity give the effective (which may be fractional when an average is taken) cluster sizes in each of the states. A value smaller than unity indicates a degeneration of the micro-system to a smaller one in a hierarchical fashion, partially if it is a fraction. In other words we are considering a scenario where clusters may form superclusters or be composed of subclusters, with a corresponding change of scale in terms of the most basic unit obtainable. We have dealt elsewhere with the interesting question an oligo-parametric hierarchical structure of complex systems \cite{FS2}, but here, we restrict ourselves to cluster hierarchy changes that do no qualitatively change the description of the system.

Hence, if we take the divergence of the vector $v_i$ in the $q_i$ space, it is a measure of the escape of systems from a given configuration of correlated clusterings. And, inversely, the negative of the divergence shows the net influx of systems into an infinitesimal cell with cluster sizes $q_i$. If all the  $q_i$ are unity, then we have unfragmented and also non-clustered, i.e. uncorrelated units at that hierarchy level.

We can argue first from the point of view of statistical mechanics that this negative divergence or influx of probability, may be interpreted as entropy.

 \begin{eqnarray}\label{ent}
 S = - \sum_i \frac {\partial p_i^{q_i}} {\partial q_i}\\ \nonumber
 = - \sum_i \log(p_i) p_i^{q_i}
 \end{eqnarray}

 We know that the free energy is defined by

\begin{equation}
A = U- T S
\end{equation}

where $A$ is the free energy, $U$ is the internal energy, and $S$ is the entropy. $T$ is the temperature,
or a measure of the average random thermal energy per unit with Boltzmann constant $k$ chosen to be unity,
and hence $S T$ is a measure of the random influx of energy into the system due to the breaking/making of
correlated clusters due to random interactions in a large system. This allows the subtracted quantity $A$
to be the free energy or the ``useful" energy. There are usual thermodynamic phase space factors in dealing with a macroscopic system, which we drop as common factors in what follows.

In terms of the Shannon coding theorem also we arrive at the same expression. Since, with $q_i$ average
clustering  the i-th state occurs with probability $p_i^{q_i}$, a stream of units (clustered) emitted will
correspond to the probability

\begin{equation}
\exp[-S] = \prod_i p_i^{p_i^{q_i}}
\end{equation}

which too gives us Eqn. \ref{ent}.

In \cite{FS1} we have developed the statistical, mechanics of this entropy in detail,
by first obtaining its probability distribution function in terms of the Lambert W function.
However, we used, for simplicity an isotropic (in state space) correlation and rescaling, i.e.
we had a single common $q$, as in Tsallis' entropy. In the rest of this work we shall use the same simpler expression.

\section{Quantum Entanglement and Stochasticity}

When the state vector in the combined Hilbert space ${\mathit H}$ of
two particles (or subsystems in ${\mathit H_A}$ and ${\mathit H_B}$)
cannot be expressed as the factorizable product of vectors in the
Hilbert spaces of the two subsystems, it is by definition entangled.
Hence entanglement is actually a property related to projection in
the subspaces, and cannot be expected to be measurable by properties
in the bigger space alone.

Given the state

\begin{equation} \label{psic}
|\psi\rangle = \sum_{ij} C_{ij}|i\rangle_A |j\rangle_B
\end{equation}

We get the density matrix for the product space

\begin{equation}\label{rhoABc}
\rho_{AB} = C_{ij} C^*_{kl}|ij\rangle_{AB} \langle kl|_{AB}
\end{equation}

which gives density matrix for the $\it{H_A}$ space for the trace
over the $\it{H_B}$ part

\begin{equation}\label{rhoAC}
\rho_A  = C C^\dag
\end{equation}

where the $C's$ are now the coefficient matrices. In terms of
density matrices an entangled state (for an explicit example) of two
qubits ( a ``qubit", or quantum bit, being a quantum superposition of two possible states) may be expressed by the reduced $2X2$ matrix from the $|0\rangle |0\rangle, |1\rangle,|1\rangle$ basis sub-set:

\begin{equation} \label{rhogam}
\rho = \left[ \begin{array}{ccc}
c^2 & \gamma c s \\
\gamma c s  &    s^2\\
\end{array} \right]
\end{equation}

with $\gamma=1$ for the pure quantum (entangled) state

\begin{equation} \label{psient}
|\psi\rangle  = c |0\rangle |0\rangle +  s |1\rangle |1\rangle
\end{equation}

and we have used the compact notation $c= \cos(\theta)$, and $s=
\sin(\theta)$. This entanglement occurs in the subspace of the
product Hilbert space involving only the two basis vectors
$|00\rangle$ and $|11\rangle$. Other entangled combinations are
equivalent to this form and may be obtained from it simply by
relabeling the basis vectors, and hence we shall use this as the
prototype.

 For $|\gamma | <1$, we have an impure state with  a classical
 stochastic component in the probability distribution, although we
 still have probability conservation as $Tr (\rho) = 1$ which remains unchanged under
 any unitary transformation. A possible measure of the  factorizability (``purity" \cite{VE1})  $\zeta$ of a quantum state, or its  quantum non-entanglement, remains invariant under changes of $\gamma$

 \begin{eqnarray} \label{zeta}
 \zeta = Tr_A [ Tr_B(\rho_{AB})^2] \nonumber \\
 = c^4 + s^4.
 \end{eqnarray}

 So, for the maximum entanglement $\zeta=1/2$ when $\theta=\pi/4$, and the
 minimal entanglement corresponds to $\zeta= 1$ (pure factorizable states) when
 $\theta=0,  \pi/2$.

Quantum impurity represented by classical stochasticity attains the maximum value when $\gamma=0$, and is nonexistent when $\gamma=1$ corresponding to a pure entangled state.

We note that $\zeta$ does not involve the stochasticity-related
parameter $\gamma$ at all, but remains the quantifier of the quantum
entanglement.

Another equivalent but conceptually possibly more interesting way of
quantifying entanglement may be the parameter

\begin{eqnarray} \label{traceAB}
E_{AB}= 2( Tr[\rho_{AB}] - Tr[\rho_A]Tr[\rho_B]) = \sin^2(2 \theta)
\end{eqnarray}

which is more symmetric in the two subspaces and is similar to a correlation function. It gives 0 for no entanglement when $\theta=0, \pi/2$, and maximal entanglement 1 for
$\theta=\pi/4$, as desired. This definition of entanglement is in
the spirit of mutual information, though we have not used the
entropy at this stage, but only the probabilities directly. It too
does not involve the stochasticity in terms of the purity parameter
$\gamma$. In the relation above we have used

\begin{equation}\label{rhoABtr}
\rho_{A}= Tr_{B} [\rho_{AB}]
\end{equation}

and similarly for $\rho_B$. In our specific case,for $A$ or for $B$,

\begin{equation}\label{rhoAmat}
\rho_{A,B} =  \left[ \begin{array}{ccc}
c^2 &  0 \\
0  &   s^2\\
\end{array} \right]
\end{equation}

with $Tr[\rho_{A,B}] = 1$ ensured.

\section {Stochasticity from Entanglement with Environment}

It is possible to formulate the stochasticity by coupling the
entangled state $|\Psi_{AB}>$ to the environment state $|\Psi_E>$
quantum mechanically and then taking the trace over the environment
states.

\begin{equation}\label{PsiABE}
|\Psi_{ABE}\rangle = \sum_{ijk} c_{ijk} |i\rangle_A |j\rangle_B
|k\rangle_E
\end{equation}

Hence the density matrix for the pure quantum system is

\begin{equation}\label{rhoABE}
\rho_{ABE} = \sum_{ijk,lmn}c_{ijk}c^*_{lmn}|ijk\rangle\langle lmn|
\end{equation}

and the trace over the environment yields

\begin{equation}\label{rhoABcc}
\rho_{AB} = \sum_{ij,lm}c_{ijk}c^*_{lmk} |ij\rangle\langle lm|
\end{equation}

For the entangled mixture of $|00\rangle$ and $|11\rangle$ in
${\mathit H_{AB}}$ and the couplings

\begin{eqnarray}\label{carr}
c_{000} = c c'  \nonumber  \\
 c_{001} = c s'  \nonumber  \\
  c_{110}= s s' \nonumber  \\
c_{111}= s c'
\end{eqnarray}

with $c'= \cos(\theta')$ and $s' = \sin(\theta')$,

the trace over $\it{H_E}$ states gives the density matrix

\begin{equation}\label{rhoc'}
 \rho_{AB}= \left[ \begin{array}{cc}
c^2 &  2 c s c's'  \\
2 c's'c s  &    s^2
\end{array} \right]
\end{equation}

So, classical stochasticity has been introduced by taking the trace
over the environment space with

\begin{equation}\label{gamma}
\gamma = \sin(2 \theta')
\end{equation}

\section{Entanglement, Entropy and Mutual Information}

\subsection{Single System Interacting with Environment}

Let us consider a single system $A$ interacting with the environment $E$. The product space ${\it H_A}\bigotimes {\it H_E}$ contains
entanglement between the measured system and the environment, and
hence the density operator for the combined system-environmental
space is as given in Eqn. \ref{rhogam} with $\gamma=1$ to indicate a
pure entangled state, and the environment-traced density is given by
Eqn. \ref{rhoABtr}. Here $\rho_A$ and $\rho_E$ are equal. Mutual
information may be identified as the entanglement, and  defined by

\begin{eqnarray}\label{mut}
E_{AE}= 2(Tr_{AE}[\rho_{AE}] - (Tr_{A}[\rho_A])^2) = \sin^2(2\theta')
\end{eqnarray}

as before, with $\theta'$ the angle of entanglement.  Hence,
measurements on the system $A$ reflects the coupling of  the system
to the environment to the environment, and the mutual information
is contained in the parameters of the system itself. In terms of
von Neumann entropy, which, with mixing in an orthogonal quantum basis, becomes similar to Shannon entropy \cite{ME1},

\begin{equation} \label{entropy}
S = -Tr [ \rho  \log (\rho) ]
\end{equation}.

We know from the Araki-Lieb relation\cite{AR1}

\begin{equation} \label{Araki}
S_{AE} \geq |S_A - S_E|
\end{equation}

that, with $S_{AE} = 0$ in a pure quantum state, we must have $S_A =
S_E$ , and hence

\begin{eqnarray} \label{Armut}
I_{AE} = S_A + S_E - S_{AE} = 2 S_A \nonumber  \\
 = -2 Tr_A [\rho_A\log(\rho_A)]
\end{eqnarray}

which too confirms the view that the system itself contains in its parameters the
mutual information in such a case, as we found above.

If we use our new form of entropy\cite{FS1} with the hypothesis
that the mutual information is still given by the same form with the
parameter $q$ not equal to $1$, which is the case for Shannon
entropy, then we get

\begin{eqnarray} \label{myinfo}
I_{AE} = - 2 Tr_A[ \rho_A^q \log(\rho_A)] \nonumber  \\
 = - 2c'^{2q}\log(c'^{2q}) -2 s'^{2q} \log(s'^{2q})
\end{eqnarray}

\subsection{Entangled Systems interacting with the Environment}

With the 3-system entanglement shown in Eqn. \ref{PsiABE} and the
relatively simple choice of couplings in Eqn. \ref{carr}, we have
already shown $\rho_{AB}$ in Eqn. \ref{rhoc'}. Similar construction
of $\rho_{AE}$,  $\rho_{BE}$ and $\rho_{ABE}$, and defining the
3-system mutual information as

\begin{eqnarray}
I_{ABE}(q) = -S_{ABE}(q)+ S_{AB}(q)+S_{BE}(q) + S_{AE}(q) \nonumber
\\ -S_A(q)-S_B(q)-S_E(q)
\end{eqnarray}

and with $S_{ABE}(q) = 0$  for any $q$, for a single 3-system pure
state, we may find the 3-system mutual information.

Tracing over the $B$ space gives $\rho_{AE}$, which, using as basis
$|00\rangle$, $|01\rangle$, $|10\rangle$ and $|11\rangle$ in the
$|AE\rangle$ product space, yields

\begin{equation}\label{rhoAE}
 \rho_{AE}= \left[ \begin{array}{cccc}
c^2c'^2 & c^2c's' & 0 & 0\\
 c^2c's' & c^2s'^2 & 0 &0\\
 0 & 0 & s^2s'^2 &  s^2c's' \\
 0  & 0 & s^2c's'
& s^2c'^2
\end{array} \right]
\end{equation}

and an identical matrix for $\rho_{BE}$.

Finally we get using the relevant eigenvalues

\begin{eqnarray}  \label{3mut}
I_{ABE}(q) =  c'^{2q} \log(c'^{2q})+ s'^{2q} \log(s'^{2q})
\nonumber \\
 -\lambda_+^{q} \log (\lambda_+^{q})-
\lambda_-^{q} \log (\lambda_-^{q})
\end{eqnarray}

where the eigenvalues $\lambda_1$ and $\lambda_2$ are for the
$\rho_{AB}$ matrix obtained after tracing over $E$-space.

\begin{equation}
\lambda_{+,-}= (1/2) ( 1 \pm {\surd[{1- 4( 1- \gamma^2) c^2s^2}}])
\end{equation}

with $\gamma$ given by Eqn. \ref{gamma}.

Had we started with a stochastic picture of entangled impure A-B
system, with $1-\gamma$ representing the stochasticity, as we have
suggested above, then the mutual information would be

\begin{eqnarray}\label{ABmut}
I_{AB}(q)  = -S_{AB}(q)+ S_A(q) + S_B(q) \nonumber  \\
 =\lambda_+^{q} \log(\lambda_+^{q})+ \lambda_-^{q} \log (\lambda_-^{q})
 \nonumber \\
 - 2 c^{2q}\log(c^{2q}) - 2 s^{2q} \log(s^{2q})
\end{eqnarray}

\begin{figure}[ht!]
\begin{center}
\includegraphics[width=8cm]{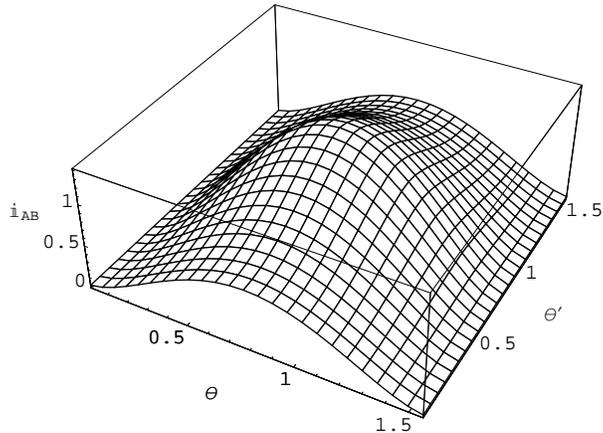}
\caption{\label{fig1}$I_{AB}$ as a function of entanglement angle
$\theta$ in A-B space and the entanglement angle $\theta'$ with the
environment, which is related to the stochasticity.}
\end{center}
\end{figure}

In Fig. \ref{fig1} we first show the mutual information (MI)
calculated according to the Shannon form of the entropy, which is
equivalent to our form at $q=1$, as a function of the $A-B$
entanglement angle $\theta$ and the entanglement angle of
$(A-B)$-system with the environment $\theta'$, which is related to the
stochasticity $\gamma$ as explained above. We note that mutual information is virtually independent of the angle of entanglement with the environment $\theta'$. Hence, it seems that traditional entropy in this case is insensitive to details of coupling with the environment when the mutual information between two systems is measured.

\begin{figure}[ht!]
\begin{center}
\includegraphics[width=8cm]{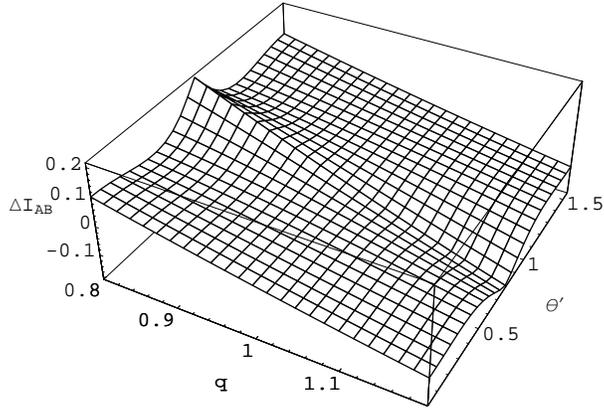}
\caption{\label{fig2}Difference of MI from our entropy with that
from Shannon entropy at $\theta=\pi/4.$}
\end{center}
\end{figure}

\begin{figure}[ht!]
\begin{center}
\includegraphics[width=8cm]{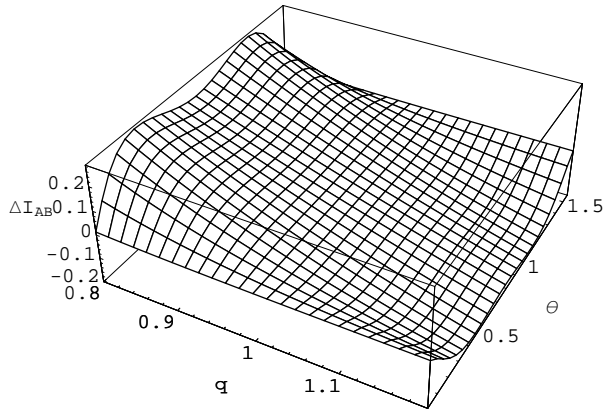}
\caption{\label{fig3}Same as Fig. \ref{fig2} with $\theta'=\pi/4$}
\end{center}
\end{figure}

In Fig. \ref{fig2} and in Fig. \ref{fig3} we show the deviations
$\Delta I_{AB}$ of our MI from the Shannon MI, as a function $q$ and
entanglement angles $\theta'$ and $\theta$ respectively, keeping the
other angle at $\pi/4$  in each case. There is symmetry around $\pi/4$. The variation is fairly smooth for fixed $\theta'$, i.e. fixed entanglement with the environment. However, if the entanglement between $A$ and $B$ is kept fixed at near $\pi/4$, then the mutual information using our form of entropy changes sharply with $\theta'$ near the symmetry value $\theta'= \pi/4$. One can see that this comes from one of the eigenvalues of the density matrix $\rho_{AB}$ approaching zero for this mixing value, and with $q \neq 1$, the there is either a sharp peak or dip compared to the Shannon entropy case, which has fixed $q =1$.

Fig. \ref{fig2} shows little variation with changing $\theta'$ for
almost any $q$. Fig. \ref{fig3} shows pronounced changes at small q
( $q < 1 $)for different $\theta$.

\begin{figure}[ht!]
\begin{center}
\includegraphics[width=8cm]{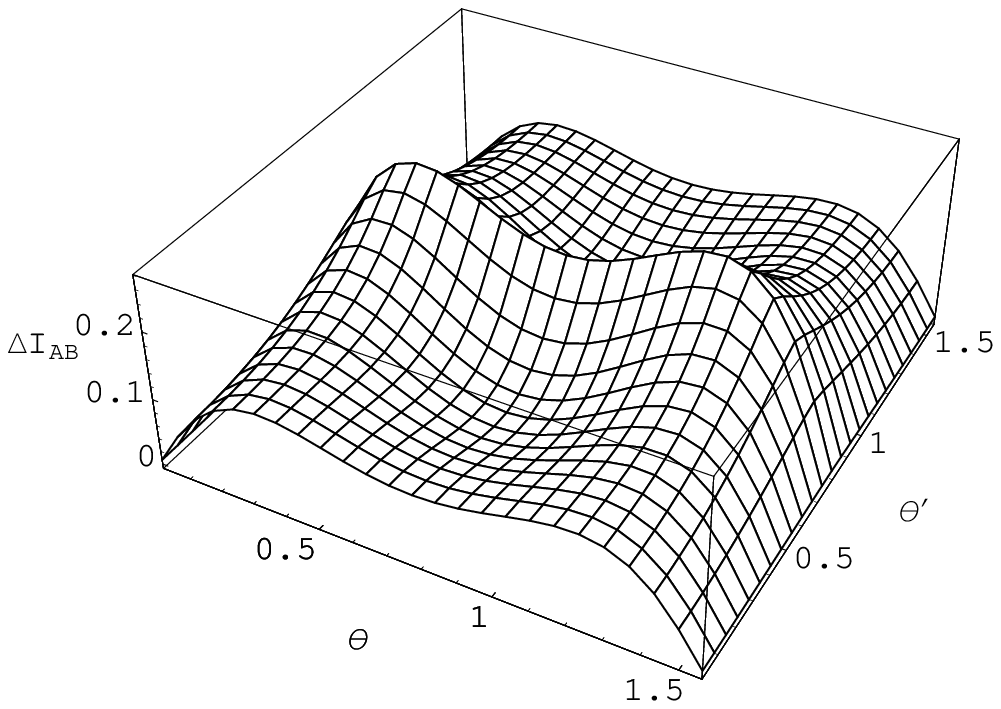}
\caption{\label{fig4}MI difference between our entropy form and
Shannon for $q=0.7$}
\end{center}
\end{figure}

\begin{figure}[ht!]
\begin{center}
\includegraphics[width=8cm]{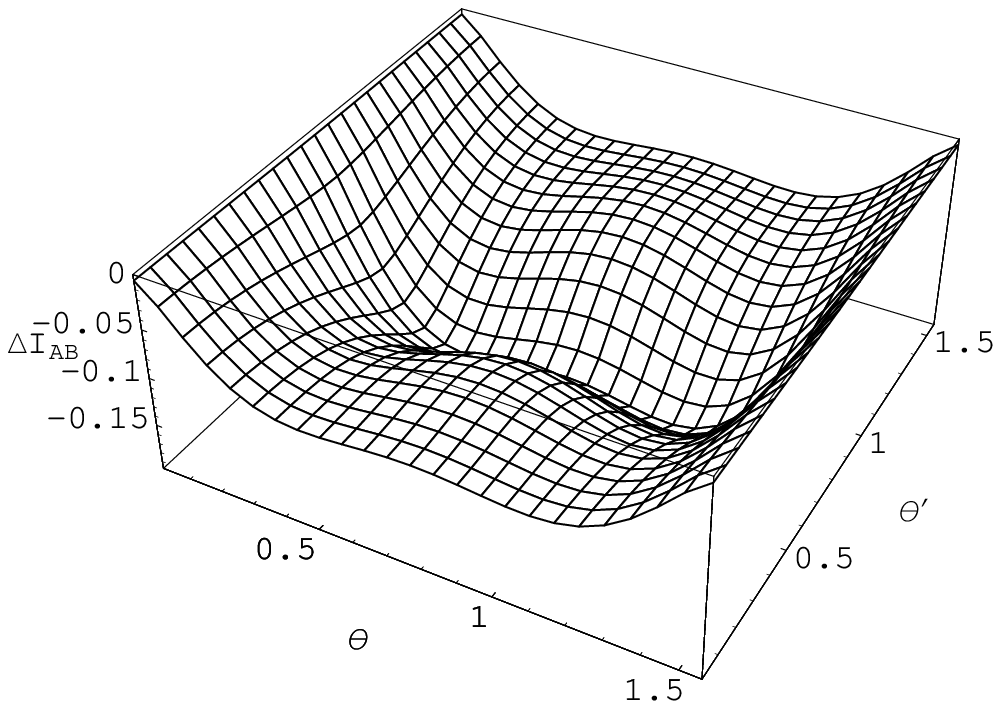}
\caption{\label{fig5}Same as Fig. \ref{fig4} but for $q=1.3$}
\end{center}
\end{figure}

In Fig. \ref{fig4} and Fig. \ref{fig5} we show the difference
between our MI and the Shannon MI as a function of $\theta$ and
$\theta'$ simultaneously, keeping $q$ fixed at $0.8$ and at $1.2$.
Of course at $q=1$, we get no difference, as our entropy then
coincides with the Shannon form. Here too we notice that the mixing
angle between $A$ and $B$ shows fairly smooth variation, but $\theta'$ or equivalently the stochasticity, causes pronounced peak ( for $q<1$), or dip (for $q>1$). Here too we can conclude that our method of entropy calculation can indicate a greater role of the entanglement with the environment when this mixing is nearly equal for the $A-B$ entangled states.

\begin{figure}[ht!]
\begin{center}
\includegraphics[width=8cm]{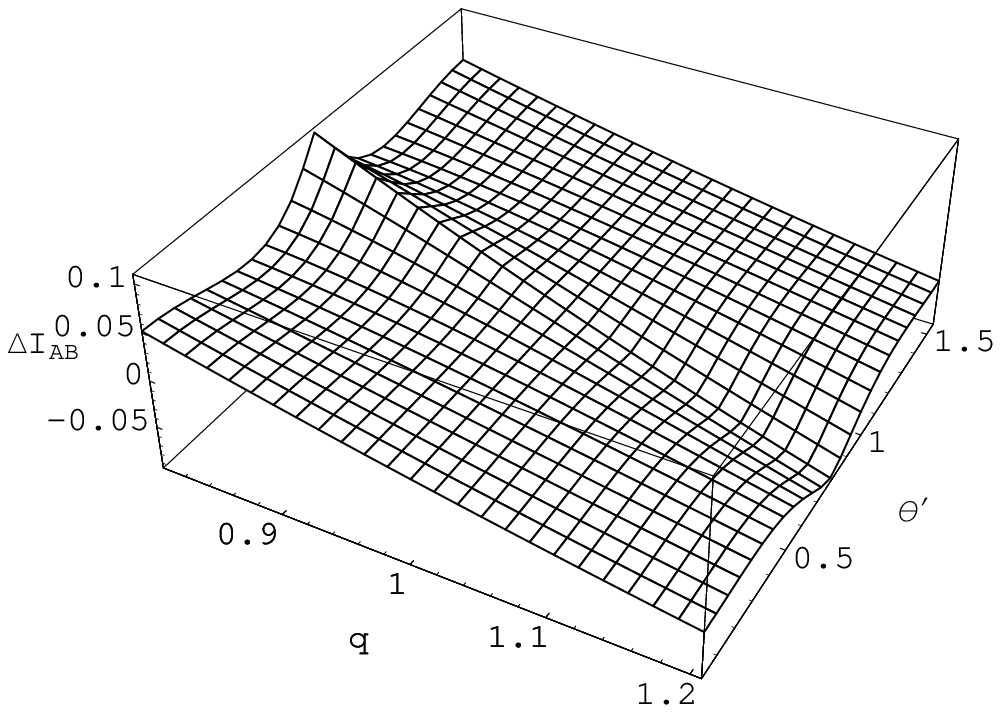}
\caption{\label{fig6}Difference between MI from our entropy and
Tsallis's with $\theta=\pi/4$}
\end{center}
\end{figure}

\begin{figure}[ht!]
\begin{center}
\includegraphics[width=8cm]{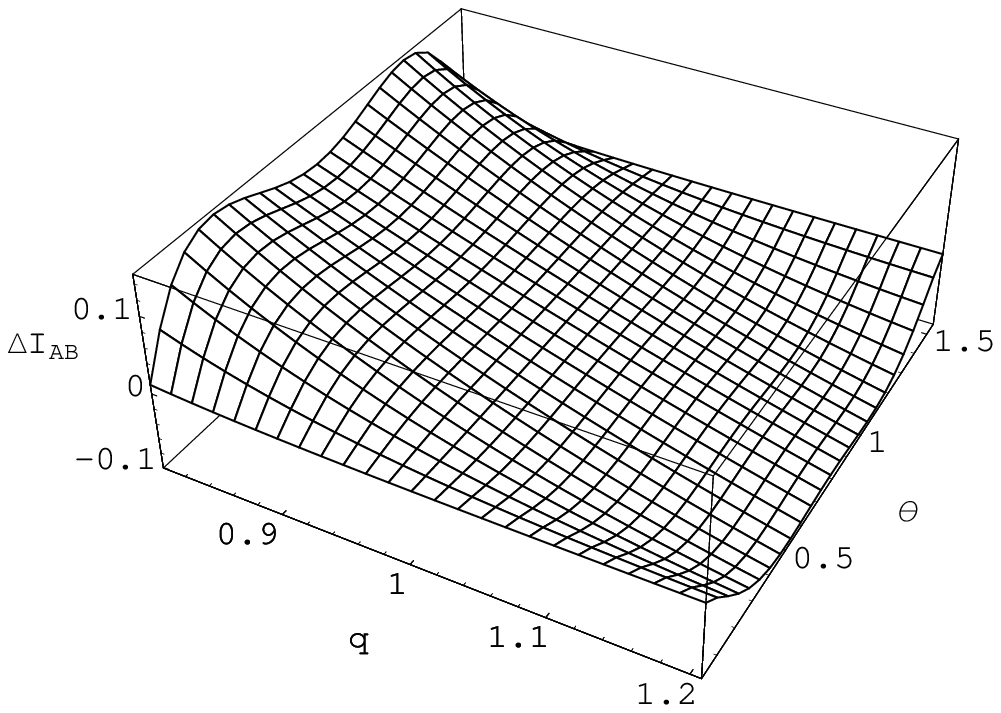}
\caption{\label{fig7}Same as Fig. \ref{fig6} but with
$\theta'=\pi/4$.}
\end{center}
\end{figure}

In view of the prevalent familiarity with the Tsallis form of
nonextensive entropy, we have previously \cite{FS1} compared our form
with results from Tsallis entropy in formulating a general
thermodynamics. There we showed that despite the conceptual and
functional differences between Tsallis entropy and our new form, the
results are very similar if we take the Tsallis $q$ to be twice as
far from unity (the Shannon equivalent value) as our value of $q$.
In Fig. \ref{fig6} and Fig. \ref{fig7} we show the difference of our
MI from that derived from Tsallis's entropy. We again note that among $\theta$
and $\theta'$ differences are both relatively more significant for $q$
values different from  $1$, for both angles near $pi/4$, with peaks and dips similar to the comparison with the mutual information calculated with Shannon entropy.

It is interesting to note here that recently in a study of the
entropy of a chain of spins in a magnetic field \cite{CH1} it has
been found that both the usual von Neumann and Renyi forms of
entropy yield nonzero and surprisingly simple closed expressions.
Though this work does not mention entanglement explicitly, the
correlation functions presented here, which determine the density
matrix and hence its diagonalized form needed for entropy
calculation, actually are manifestations of the entanglement among
the spins and between the spins and the magnetic field. The chain
has been split into two parts, similar to our A and B subsystems,
and the external magnetic field acts like the environment we have
introduced in this work. Though they carry out their extensive
calculations at zero temperature, unlike our finite temperature
treatment, the fact they obtain nonzero $S_A$ for the first $L$
spins is apparently due to the segmentation of the pure state of the
fully entangled quantum system and the consideration of part $A$
only for the entropy calculation,  which effectively is equivalent
to summing the states of part $B$ and the entanglement with the
environment, and produces entropy due to the corresponding loss of
information about state of the whole system. Hence, their results
for this explicit model is consistent with our general result that
classical stochasticity and entropy may be a reflection of segmented
consideration of bigger complete systems. The values of the entropy
of different types such as the canonical Shannon form or generalized
forms such as the Renyi form, which goes to the Shannon form in the
usual limit, like that of the related parameter we have mentioned
for Tsallis entropy and for our new form of entropy in this work,
reflect the extent of entanglement or interaction or, equivalently,
correlation.  In their work a length scale comes out of this
segmentation, which appears to be similar to the angle of
entanglement in our case. We do not get a phase transition as they
do, because we have considered a simplified general finite system of
only two or three component subsystems, not an infinite chain, and
finite systems cannot show any phase transitions.

\section{Conclusions}

We have shown how a simple definition of the entropy in terms of influx of states into cells of any given cluster sizes in various states can give us a new nonextensive form of entropy with a closed form of the probability distribution function, and which coincides with the usual form for uncorrelated microsystems. We have then seen that classical stochasticity can be derived from quantum entanglement with the environment, and it influences the mutual information between quantum states. The functional form of
the MI differs as per the definition of the entropy, but the
numerical differences in MI resulting from various forms of entropy
usually differ rather subtly according the parameters of entanglement, within
the system, and with the environment, as well as the scaling type
parameter $q$ introduced in the Tsallis form and in the new form
introduced by us. However, for the angle of entanglement with the environment our entropy differs from both the Shannon entropy and Tsallis entropy when both angles of entanglement   are near the symmetry point $\pi/4$. The differences between the forms become more
pronounced as $q$  varies from unity. Entanglement and mutual information are
such fundamental concepts that experimental tests need to be
designed to distinguish from minute quantitative differences the
appropriateness of various theoretical forms of entropy. Theoretical
works like the study of large systems such as spin chains \cite{CH1}
may also help differentiate the appropriateness of various forms of
entropy such as Shannon, Tsallis, Renyi or our suggested new form,
and the quantification of mutual information. The notion of clusters changing constantly into various sizes, which is the basis of our definition of the new form of generalized entropy may be the most relevant concept for the treatment of liquids and other material, where such phenomena form an integral part of the dynamics.

\section*{Acknowledgements}

The author would like to thank Andrew Tan and Ignacio Sola for
discussions and G. Kaniadakis and J.F. Collet for useful feedback.

\end{document}